\documentclass[amsmath,amssymb, 12pt]{revtex4}
\usepackage{graphicx}
\usepackage{dcolumn}
\usepackage{bm}
\usepackage{graphicx}
\usepackage{epsfig}

\def\bea{\begin{eqnarray}}
\def\eea{\end{eqnarray}}

\begin{document}
\title{The coincidence problem in the scenario of dark energy interacting with two fluids}
\author{Norman Cruz}
\altaffiliation{ncruz@lauca.usach.cl} \affiliation{Departamento de
F\'\i sica, Facultad de Ciencia, Universidad de Santiago, Casilla
307, Santiago, Chile.}
\author{Samuel Lepe}
\altaffiliation{slepe@ucv.cl} \affiliation{Instituto de F\'\i
sica, Facultad de Ciencias, Pontificia Universidad Cat\'olica de
Valpara\'\i so, Casilla 4059, Valpara\'\i so, Chile.}
\author{Francisco Pe\~na}
\altaffiliation{fcampos@ufro.cl} \affiliation{Departamento de
Ciencias F\'\i sicas, Facultad de Ingenier\'\i a, Ciencias y
Administraci\'on, Universidad de La Frontera, Avda. Francisco
Salazar 01145, Casilla 54-D Temuco, Chile.\\}
\date{\today}
\begin{abstract}
A cosmological model of dark energy interacting with dark matter
and another general component of the universe is considered. The
evolution equations for coincidence parameters $r$ and $s$, which
represent the ratios between the dark energy and the matter and
the other cosmic fluid, respectively, are analyzed in terms of the
stability of stationary solutions. The obtained general results
allow to shed some light on the coincidence problem and in the
equations of state of the three interacting fluids, due to the
constraints imposes by the stability of the solutions. We found
that for an interaction proportional to the sum of the DE density
and the third fluid density, the hypothetical fluid must have
positive pressure, which leads naturally to a cosmological
scenario with radiation, unparticle or even some form of warm DM
as the third interacting fluid.

%\pacs{}
\end{abstract}
\maketitle
\section{ Introduction}

The existence of a dark component with an exotic equation of
state, i. e., with a  ratio $w =  p/\rho$ negative and close to
$-1$, which drives an accelerated expansion is consistent with the
luminosity distance as a function of redshift of distant
supernovae~\cite{Riess}, the structure formation
(LSS)~\cite{Percival} and the cosmic microwave background
(CMB)~\cite{Spergel}

The cosmic observations show that densities of dark energy and
dark matter are of the same order today. To solve this coincidence
problem~\cite{Weinberg} (or why we are accelerating in the current
epoch due that the vacuum and dust energy density are of the same
order today ?) it is assumed an evolving dark energy field with a
non-gravitational interaction with matter~\cite{Amendola} (decay
of dark energy to matter).

Although the main topic of investigation have been centered in the
interactions in the dark sector, it is physically reasonable and
even expected from a theoretical point of view, that dark
components can interact with other fluids of the universe. For
example DE interacting with neutrinos was investigated
in~\cite{Brokfield}, and decaying into the fermion fields,
in~\cite{Macorra}. A more general scenario was considered
in~\cite{Kremer}, in which DE is interacting with neutrinos and
DM.

Inspired in these previous investigation we have recently
formulated a effective model where DE is decaying in DM and
another hypothetical fluid~\cite{CruzLettB}, which we shall denote
here and after by DX. In the framework of the holographic DE, and
using the Hubble radius as infrared cutoff, we have shown that our
scenario leads naturally, for a flat universe, to a more suitable
approach to the cosmic coincidence problem in which the ratio
between the energy densities of DM and DE, $r$, can be variable
during the cosmic evolution. Our model has been discussed has a
possible approach to solve the triple coincidence problem
in~\cite{Jamil}. In this work it was assumed that the third fluid
is radiation. Although matter and radiation are almost
non-interacting fluids, since the decoupling era, nevertheless
they could interact with DE. In~\cite{Chen} was investigated the
dynamical behavior when DE is coupling to DM and unparticle in the
flat FRW cosmology.

The aim of this paper is to investigate further the model of DE
interacting with DM and another hypothetical fluid. Despite of the
interesting results found in~\cite{Jamil,Chen} when this third
fluid was specifically identified with radiation and unparticle,
we continue assuming only that this fluid has an equation of state
with $\omega$ constant. We expect, studying the stationary
solutions of the evolution equation for $r$ and $s$, the radio
between the energy densities of DE and DX,  to found suitable
constraints for the equation of state of the unknown fluid.  Since
we are introducing a third fluid in the interactions in the dark
sector, we generalize the coupling terms that have been already
considered in the literature, although and interaction term which
include the energy density of unparticle was discussed
in~\cite{Chen}.  In this paper we choose three different coupling
terms in order to investigate the dynamical behavior of the models
of DE interacting with DM and a third unknown fluid.

Our paper is organized as follows. In section II we present the
model for a universe filled with dark matter, dark energy and
another fluid. We shall impose that the interacting term $\delta$
and $\delta'$ which appears in the conservation equations are
different. In section III we study the stationary solutions and
its stability for the equations of evolution of the ratios between
the DM and DE and the third fluid and DE. We presents the
constraints on the equation of state for DE and the hypothetical
third fluid. In section IV we discuss our results.

\section{Interacting dark energy}

In the following we modeled the universe made of CDM with a
density $\rho_{m}$ and a dark energy component, $\rho_{D}$. We
will assume that the the dark matter component is interacting with
the dark energy component, so their continuity equations take the
form
\begin{eqnarray}\label{continuity1}
\overset{\cdot }{\rho }_{D}+3H\left( 1+\omega_{D} \right)\rho _{D}
&=&-\delta^{\prime},
\end{eqnarray}
\begin{eqnarray}\label{continuity2}
\overset{\cdot }{\rho }_{m}+3H\left( 1+\omega_{m} \right)\rho
_{m} &=&\delta.
\end{eqnarray}
where $\delta$ and $\delta^{\prime}$ are the interactions terms,
which are different in order to include the scenario in which the
mutual interaction between the two principal components of the
universe leads to some loss in other forms of cosmic constituents.
We assume that the interactions terms are of the form $\delta
=3H\Gamma$, where $\Gamma=\Gamma (\rho _{D},\rho _{m}, \rho
_{X})$. Since we will study an universe with dark energy decaying
into dark matter, we have from the beginning $\delta >0$ and
$\delta^{\prime}>0$. If we denote this other component by
$\rho_{X}$, its corresponding continuity equation is given by
\begin{eqnarray}\label{continuity3}
\overset{\cdot }{\rho }_{X}+3H\left( 1+\omega_{X} \right)\rho _{X}
&=&\delta^{\prime}-\delta= 3H\,\Pi,
\end{eqnarray}
where $\Pi =\Gamma^{\prime}-\Gamma$. We are taking about in this
case that dark energy decay into dark matter (or viceversa,
depending on the sign of $\delta$) and other component. The
sourced Friedmann equation is then given by
\begin{eqnarray}\label{sourcedFridman}
3H^{2} &=&\rho _{D}+\rho _{m}+\rho_{X}-\frac{3k}{a^2}.
\end{eqnarray}
Since we are interested in the three cosmic coincidence, we do not
we assume as it was done in~\cite{CruzLettB}, that $\rho _{D}\gg
\rho_{X}$ and $\rho _{m}\gg \rho_{X}$.  In order to study the
evolution of the densities of these three fluids we construct the
differential equations for the coincidence parameters $r=\rho
_{m}/\rho _{D}$ and $s=\rho _{X}/\rho _{D}$. These equations takes
the following expressions
\begin{eqnarray}\label{requation}
r^{\prime}= \frac{\dot{r}}{H} &=& \frac{r}{H} \left(
\frac{\dot{\rho_{m}}}{\rho_{m}} - \frac{\dot{\rho_{D}}}{\rho_{D}}
\right),
\end{eqnarray}
and
\begin{eqnarray}\label{sequation}
s^{\prime}= \frac{\dot{s}}{H} &=& \frac{s}{H} \left(
\frac{\dot{\rho_{X}}}{\rho_{X}} - \frac{\dot{\rho_{D}}}{\rho_{D}}
\right) .
\end{eqnarray}
We assume that the interaction term takes the general form
\begin{eqnarray}\label{interaction}
\Gamma = \, \lambda_{1} \rho_{D}+ \lambda_{2}\rho_{m}+
\lambda_{3}\rho_{X} = \, \left( \lambda_{1} + \lambda_{2} r+
\lambda_{3}s \right) \rho_{D} .
\end{eqnarray}
Consistently, for the general form given in
Eq.(\ref{interaction}), we can parameterized $\Pi
=\Gamma^{\prime}-\Gamma$ throughout suitable constants
$\lambda_{1}^{\pi}, \lambda_{2}^{\pi},\lambda_{3}^{\pi}$, in the
following form
\begin{eqnarray}\label{Piinteraction}
\Pi = \, \lambda_{1}^{\pi} \rho_{D}+ \lambda_{2}^{\pi}\rho_{m}+
\lambda_{3}^{\pi}\rho_{X}=  \,  \left( \lambda_{1}^{\pi} +
\lambda_{2}^{\pi} r+ \lambda_{3}^{\pi}s \right) \rho_{D} .
\end{eqnarray}
Notice that the form of interaction considered here is a
generalization of the cases previously investigated, since the
interactions taken account consider only functions of the dark
sector densities. Interactions that are linear combinations of the
dark sector densities have been studied in~\cite{Interactions},
when the interaction is only between the dark components.

Using Eqs.(\ref{continuity1}) and (\ref{continuity2}) in the
equations (\ref{requation}) and (\ref{sequation}), with the
interaction terms, $\Gamma$ and $\Pi$ given by Eqs.
(\ref{interaction}) and (\ref{Piinteraction}), respectively, we
obtain the evolution equations for the parameters $r$ and $s$
\begin{eqnarray}\label{requationprime}
r^{\prime}=  3r \left[ ( \lambda_{1} + \lambda_{2} r+ \lambda_{3}s
)(1+1/r) + ( \lambda_{1}^{\pi} + \lambda_{2}^{\pi} r+
\lambda_{3}^{\pi}s ) + \omega_{D}-\omega_{m}  \right],
\end{eqnarray}
and
\begin{eqnarray}\label{sequationprime}
s^{\prime}=  3s \left[ \lambda_{1} + \lambda_{2} r+ \lambda_{3}s
+(1+1/s)(\lambda_{1}^{\pi} + \lambda_{2}^{\pi} r+
\lambda_{3}^{\pi}s) + \omega_{D}-\omega_{X} \right ].
\end{eqnarray}

\section{Stationary solutions}

Before to look for the stationary solutions of
Eqs.(\ref{requation}) and (\ref{sequation}), let us briefly
discuss the what we assume for the equations of state of the three
cosmic fluids that are under interaction.  At this stage, we only
consider that one fluid, with an energy density $\rho_{D}$ and
equation of state $\omega_{D}$, is decaying into the other two
fluids. Although, it is reasonable to take $\omega_{m}=0$ from the
beginning, since we are thinking in the dark matter fluid, we
shall postpone this election until to obtain the constraint
derived from the study of the stationary solutions of
Eqs.(\ref{requation}) and (\ref{sequation}) and its stability. For
simplicity, the three equations of state are taken constant.

Since we are looking for stationary solutions of
Eqs.(\ref{requation}) and (\ref{sequation}) we set $r^{\prime}=
s^{\prime}=0$, obtaining a systems of algebraic equations in
terms of the the variables $r$ and $s$, with the parameters
$\lambda_{1}, \lambda_{2}, \lambda_{3},\lambda_{1}^{\pi},
\lambda_{2}^{\pi}, \lambda_{3}^{\pi}, \omega_{D},\omega_{X}$. In
order to study how suitable interactions leads to these three
cosmic fluids to have stationary solutions we begin with the
simplest cases.

%------------------------------------------------------------------
%------------------------------------------------------------------
%------------------------------------------------------------------
%  CASO  CON  GAMMA=  \lambda DE
%------------------------------------------------------------------
%------------------------------------------------------------------
%------------------------------------------------------------------

\subsection{Case $\Gamma =  \lambda \rho_{D}$ and $\Pi = \lambda^{\pi} \rho_{D}$}

In this case we are choosing $\lambda_{2}= \lambda_{3}=
\lambda_{2}^{\pi}= \lambda_{3}^{\pi}=0$. The interactions terms
are proportional to the dark energy density. This type of
interaction has been investigated in~\cite{delCampo,Chen,He}. The
condition $r^{\prime}= s^{\prime}=0$ leads to the algebraic
systems
\begin{eqnarray}\label{rcero}
f(r,s)|_{r=r_{s},s=s_{s}}= \,\,\lambda(1+r_{s}) +
\lambda^{\pi}r_{s}+ (\omega_{D}-\omega_{m})r_{s}=0.
\end{eqnarray}
\begin{eqnarray}\label{scero}
g(r,s)|_{r=r_{s},s=s_{s}}=\,\,\lambda s_{s}+
\lambda^{\pi}(1+s_{s})+(\omega_{D}-\omega_{X})s_{s}=0.
\end{eqnarray}
Notice that the interaction assumed gives two simple linear
equation non acopled in the variables $r_{s}$ and $s_{s}$, which
are the stationary solutions given by
\begin{eqnarray}\label{rcerosol}
r_{s}=- \frac{\lambda}{\lambda
+\lambda^{\pi}+\omega_{D}-\omega_{m}},
\end{eqnarray}
and
\begin{eqnarray}\label{scerosol}
s_{s}=- \frac{\lambda^{\pi}}{\lambda
+\lambda^{\pi}+\omega_{D}-\omega_{X}}.
\end{eqnarray}
Since $r_{s}$ and $s_{s}$ are positive quantities, the
denominators of the above both equations must be negative, so be
obtain the following inequalities
\begin{eqnarray}\label{inqomega}
\omega_{D}<\omega_{m}- (\lambda +\lambda^{\pi}),
\end{eqnarray}
and
\begin{eqnarray}\label{inqomegax}
\omega_{X}> \omega_{D}+ (\lambda +\lambda^{\pi}).
\end{eqnarray}

In order to study the stability of the found solutions we shall
evaluate the eigenvalues of the matrix $\mathbf{M}$
\begin{eqnarray}\label{matrix}
\mathbf{M}=
\begin{pmatrix}
 _{\frac{\partial f(r,s)}{\partial r}} & _{\frac{\partial f(r,s)}{\partial s}} \\
  _{\frac{\partial g(r,s)}{\partial r}} & _{\frac{\partial g(r,s)}{\partial s}}
\end{pmatrix}
\end{eqnarray}
whose elements are evaluate at the critical point $(r_{s},s_{s})$.
From the equation for the eigenvalues, $det [\mathbf{M}-\eta
\mathbf{I}]$, and since $\partial f(r,s)/\partial s = \partial
g(r,s)/\partial r =0$, we obtain
\begin{eqnarray}\label{eqeingenvalue}
\left [\left (\frac{\partial f(r,s)}{\partial r}\right
)_{|_{r=r_{s},s=s_{s}}} - \eta\right ]\left [\left (\frac{\partial
g(r,s)}{\partial s}\right )_{|_{r=r_{s},s=s_{s}}}-\eta\right ]=0.
\end{eqnarray}
The eigenvalues are then
\begin{eqnarray}\label{eingenvalue1}
\eta_{1}=  \left( \frac{\partial f(r,s)}{\partial r}\right)
_{|_{r=r_{s},s=s_{s}}}= \,\,\lambda
+\lambda^{\pi}+\omega_{D}-\omega_{m}
\end{eqnarray}
and
\begin{eqnarray}\label{eingenvalue2}
\eta_{2}=\left (\frac{\partial g(r,s)}{\partial s}\right
)_{|_{r=r_{s},s=s_{s}}}=\,\,\lambda
+\lambda^{\pi}+\omega_{D}-\omega_{X}
\end{eqnarray}
The condition of stability, $\eta_{1}<0$ and $\eta_{2}<0$ is the
same just contained in Eqs.(\ref{inqomega}) and (\ref{inqomegax}).
So, if for a given $\omega_{D}$, $\omega_{m}$, $\omega_{X}$,
$\lambda$, $\lambda^{\pi}$ there are positive stationary
solutions, they are also stable. For the case of dark matter with
negligible pressure, i.e., $\omega_{m}=0$, Eq.(\ref{inqomega})
implies that the dark energy must necessarily have an equation of
state with $\omega_{D}< 0$. A rough estimation of the value of the
sum $\lambda +\lambda^{\pi}$ can be obtained from the equation for
the acceleration
\begin{eqnarray}\label{acceleration}
\frac{\ddot{a}}{a} = -\frac{1}{6}(1+3\omega)\rho,
\end{eqnarray}
where $\rho \equiv \rho _{D}(1+r+s)$ and $\omega \equiv
(\omega_{D}+\omega_{m}r+ \omega_{X}s)/ 1+r+s$.  If an accelerated
phase is demanded we obtain the following inequality
\begin{eqnarray}\label{inqomegaequivalent}
\omega_{D}<-\frac{1}{3}\left[ 1+ (1 + 3\omega_{m})r + (1 +
3\omega_{X})s \right ].
\end{eqnarray}
In terms of the parameters $\Omega_{D}$ and $\Omega_{X}$ (for
$\omega_{m}=0$) we obtain
\begin{eqnarray}\label{inqomegaaprox}
\omega_{D}<-\frac{1}{3\Omega_{D}}( 1 + 3\omega_{X}\Omega_{X}).
\end{eqnarray}
If it is assumed that $\Omega_{X} \ll 1$, as we expect for today
for any other fluid different from the dark sector. In order to
obtain a rough estimation of the term $\lambda +\lambda^{\pi}$, we
equate the right hand side of the expressions (\ref{inqomega}) and
(\ref{inqomegaaprox}), and taking $\Omega_{D}=0.7$ we obtain
\begin{eqnarray}\label{inqomegad}
\lambda +\lambda^{\pi}\simeq 0.5,
\end{eqnarray}
so for $\omega_{D}\lesssim -0.5$ the three interacting fluids
leads to stationary and stable solutions for $r$ and $s$. From the
inequalities (\ref{inqomega}) and (\ref{inqomegax}) we obtain that
$\omega_{X}\lessgtr \omega_{m}$. So for $\omega_{m}=0$ the third
fluid could be a normal fluid or even an exotic fluid.
Nevertheless, from the expression for the ratio
$\frac{s_{s}}{r_{s}}$, given by
\begin{eqnarray}\label{ratiors}
\frac{s_{s}}{r_{s}}= \frac{\lambda^{\pi}}{\lambda}\left
(1-\frac{\omega_{X}}{\omega_{D}+ \lambda +\lambda^{\pi}}
\right)^{-1},
\end{eqnarray}
we can conclude, assuming $\frac{s_{s}}{r_{s}}<<1$, that an
scenario with $\omega_{X}>0$ is more suitable taking
$\lambda>>\lambda^{\pi}$. From Eq. (\ref{rcerosol}) and taking
$r_{s}\approx 0.3/0.7$, $\omega_{D} \simeq -1$, we obtains for the
realistic case $\lambda>>\lambda^{\pi}$ that $\lambda \simeq 0.3$.

%------------------------------------------------------------------
%------------------------------------------------------------------
%------------------------------------------------------------------
%  CASO  CON  GAMMA=  \lambda rho m
%------------------------------------------------------------------
%------------------------------------------------------------------
%------------------------------------------------------------------

\subsection{Case $\Gamma =  \lambda \rho_{m}$ and $\Pi = \lambda^{\pi} \rho_{m}$}

In this case we are choosing $\lambda_{1}= \lambda_{3}=
\lambda_{1}^{\pi}= \lambda_{3}^{\pi}=0$. The interactions terms
are proportional only to the dark matter density. This type of
interaction was investigated for models of interacting phantom
dark energy with dark matter~\cite{Majerotto,Chen,Cai,Xchen} and
also in~\cite{XMchen,Bohmer,He}. Observational constraints on
$\lambda$ for this type of interaction have been investigated
in~\cite{Guo}. The condition $r^{\prime}= s^{\prime}=0$ leads to
the algebraic systems
\begin{eqnarray}\label{rcero1}
f(r,s)|_{r=r_{s},s=s_{s}}= \,\,r^{2}_{s} (\lambda + \lambda^{\pi})
+ r_{s}(\lambda + \omega_{D}-\omega_{m})=0.
\end{eqnarray}
\begin{eqnarray}\label{scero1}
g(r,s)|_{r=r_{s},s=s_{s}}=\,\,s_{s}r_{s}(\lambda + \lambda^{\pi})+
 \lambda^{\pi}r_{s}+(\omega_{D}-\omega_{X})s_{s}=0.
\end{eqnarray}
In this case the interaction assumed gives two coupled non linear
equations in the variables $r_{s}$ and $s_{s}$. The Eq.
(\ref{rcero1}) has the following solution different from zero ,
\begin{eqnarray}\label{rcerosol1}
r_{s}=- \frac{\lambda+\omega_{D}-\omega_{m}}{\lambda
+\lambda^{\pi}}.
\end{eqnarray}
Imposing the condition $r_{s}>0$ we obtain the constraint
\begin{eqnarray}\label{const11}
\lambda+\omega_{D}-\omega_{m} <0.
\end{eqnarray}
Introducing the value for $r_{s}$, given by Eq.(\ref{rcerosol1}),
in the Eq.(\ref{scero1}) yields
\begin{eqnarray}\label{scerosol1}
s_{s}= \frac{\lambda^{\pi}}{\lambda +\omega_{m}-\omega_{X}} \left
(\frac{\lambda+\omega_{D}-\omega_{m}}{\lambda +\lambda^{\pi}}
\right ).
\end{eqnarray}
Using the constrain given in Eq.(\ref{const11}) in the expression
for $s_{s}$ we obtain that $s_{s}>0$ implies
\begin{eqnarray}\label{inqomega1}
- \lambda + \omega_{m}-\omega_{X}< 0.
\end{eqnarray}

From the equation for the eigenvalues, $det [\mathbf{M}-\eta
\mathbf{I}]$, and since $\partial f(r,s)/\partial s = 0$, we
obtain
\begin{eqnarray}\label{eqeingenvalue}
\left [\left (\frac{\partial f(r,s)}{\partial r}\right
)_{|_{r=r_{s},s=s_{s}}} - \eta\right ]\left [\left (\frac{\partial
g(r,s)}{\partial s}\right )_{|_{r=r_{s},s=s_{s}}}-\eta\right ]=0.
\end{eqnarray}
The eigenvalues are then
\begin{eqnarray}\label{eingenvalue1}
\eta_{1}=  \left( \frac{\partial f(r,s)}{\partial r}\right)
_{|_{r=r_{s},s=s_{s}}}= \,\, 2r_{s}(\lambda +\lambda^{\pi})
(\lambda+\omega_{D}-\omega_{m})
\end{eqnarray}
and
\begin{eqnarray}\label{eingenvalue2}
\eta_{2}=\left (\frac{\partial g(r,s)}{\partial s}\right
)_{|_{r=r_{s},s=s_{s}}}=\,\,r_{s}(\lambda +\lambda^{\pi})
+\omega_{D}-\omega_{X}
\end{eqnarray}
Notice that the condition of stability, $\eta_{1}<0$ and
$\eta_{2}<0$ gives us, for $\eta_{1}<0$ the following constraint
\begin{eqnarray}\label{const22}
\lambda+\omega_{D}-\omega_{m} >0,
\end{eqnarray}
which can not be allowed if Eq.(\ref{const11}) is satisfied. Then
for this type of interaction it is not possible to have stable
solutions for three cosmic interacting fluids. Nevertheless, this
situation is a consequence of chose $\lambda >0$ and $
\lambda^{\pi}>0$ from the beginning. It is straightforward to
prove that for $\lambda +\lambda^{\pi}<0$ and $\lambda^{\pi}>0$,
or equivalently $\lambda <0$ and $|\lambda|>|\lambda^{\pi}|$, the
fixed point of the system are stable. The constraints for the
equations of state are the following
\begin{eqnarray}\label{const23}
\lambda+\omega_{D}-\omega_{m} >0,
\end{eqnarray}
and
\begin{eqnarray}\label{const24}
-\lambda+\omega_{D}-\omega_{X} <0,
\end{eqnarray}
which for $\omega_{m} =0$ yields $\lambda+\omega_{D}>0$ and
$\lambda+\omega_{X}>0$. Since $\lambda <0$ the above conditions
implies a third fluid with $\omega_{X}>0$ and also a DE with
$\omega_{D}>0$.  So in the approach of DE interacting with two
fluids, this kind of coupling give stable solutions but a cosmic
evolution without acceleration.

%----------------------------------------------------------------------------------
%  CASO  CON  GAMMA=  DE+ X
%----------------------------------------------------------------------------------

\subsection{Case $\Gamma =  \lambda \rho_{D}(1+s)$ and $\Pi = \lambda^{\pi} \rho_{D}(1+s)$}

In this case we have taken $\lambda_{1}= \lambda_{3}$,
$\lambda_{2}=\lambda_{2}^{\pi}=0$ , and $\lambda_{1}^{\pi}=
\lambda_{3}^{\pi}$.  The interaction is, in this case,
proportional to $\rho_{D}+\rho_{X}$.  A coupling term which
include a different fluid from those of the dark sector was
already introduced in~\cite{Chen}, but throughout expressions like
$\rho_{D}\rho_{X}$ and $\rho_{m}\rho_{X}$, where DX was identified
with unparticle. The condition $r^{\prime}= s^{\prime}=0$ leads to
the following algebraic system
\begin{eqnarray}\label{rcero2}
f(r,s)|_{r=r_{s},s=s_{s}}=\lambda(1+r_{s})(1+s_{s}) +
\lambda^{\pi}r_{s}(1+s_{s})+ (\omega_{D}-\omega_{m})r_{s}=0.
\end{eqnarray}
\begin{eqnarray}\label{scero2}
g(r,s)|_{r=r_{s},s=s_{s}}= \lambda s_{s}(1+s_{s})+
\lambda^{\pi}(1+s_{s})^{2}+(\omega_{D}-\omega_{X})s_{s}=0.
\end{eqnarray}
Solving first the Eq.(\ref{scero2}), which is a second grade
equation of the form $x^{2}+Bx+C=0$, where the coefficients $B$
and $C$ are given by
\begin{eqnarray}\label{scerosol2}
B=\frac{\lambda +2\lambda^{\pi}+\omega_{D}-\omega_{X}}{\lambda
+\lambda^{\pi}}; \,\,\,\,\,\,\,\,\,
C=\frac{\lambda^{\pi}}{\lambda +\lambda^{\pi}}.
\end{eqnarray}
The solutions of Eq.(\ref{scero2}) has the form
\begin{eqnarray}\label{solsecondeq}
s_{s}= \frac{B}{2}\left (-1 \pm \sqrt{1-\frac{4C}{B^{2}}}\right ).
\end{eqnarray}
Since $s_{s}$ is a positive and real number we need to impose the
two constraints $B<0$ and $B^{2}>4C$. The first one implies that
$\lambda +2\lambda^{\pi}+\omega_{D}-\omega_{X}<0$ and the second
one, $\lambda
+2\lambda^{\pi}+\omega_{D}-\omega_{X}>-2\sqrt{\lambda^{\pi}(\lambda
+\lambda^{\pi})}$, which gives the following range for
$\omega_{D}-\omega_{X}$
\begin{eqnarray}\label{inqforomegas}
- \left ((\lambda +2\lambda^{\pi})+2\sqrt{\lambda^{\pi}(\lambda
+\lambda^{\pi})}\right) <\omega_{D}-\omega_{X} <- (\lambda
+2\lambda^{\pi}).
\end{eqnarray}
Introducing the two solutions of Eq.(\ref{scero2}), which we
denote by $s_{s+}$ and $s_{s-}$, in Eq.(\ref{rcero2}), we obtain
two solutions for $r_{s}$, $r_{s+}$ and $r_{s-}$, given by
\begin{eqnarray}\label{rcerosol2}
r_{s+}= -\frac{\lambda}{\lambda +\lambda^{\pi}+
\frac{\omega_{D}-\omega_{m}}{1+s_{s+}}}; \,\,\,\,\,\,\,\,\,
r_{s-}=- \frac{\lambda}{\lambda +\lambda^{\pi}+
\frac{\omega_{D}-\omega_{m}}{1+s_{s-}}}
\end{eqnarray}
Since $r_{s}>0$ Eq.(\ref{rcerosol2}) gives the following
constraint
\begin{eqnarray}\label{inqrsol}
(\lambda +\lambda^{\pi})(1+s_{s\pm})+\omega_{D}-\omega_{m}<0 ,
\end{eqnarray}
so $\omega_{D}$ must satisfy
\begin{eqnarray}\label{inqomegad1}
\omega_{D}< \omega_{m}- (\lambda +\lambda^{\pi})(1+s_{s\pm}).
\end{eqnarray}
The conditions which are necessary to hold in order to have
positive solutions for $r_{s\pm}$ and  $s_{s\pm}$ are then the
inequalities given by Eqs. (\ref{inqforomegas}) and
(\ref{inqrsol})

As in the case \textbf{A}, where the interaction is proportional
only to the DE density, if $\omega_{m}=0$ the above inequality
implies that the dark energy must necessarily have an equation of
state with $\omega_{D}< 0$. A rough estimation for the range of
the values that the parameters $\lambda$ and $\lambda^{\pi}$ can
take, may also be done for this case using Eq.
(\ref{inqomegaaprox}) with $\Omega_{X} \ll 1$ and
$\Omega_{D}=0.7$. Equating the right hand side of the expressions
(\ref{inqomegad1}) and (\ref{inqomegaaprox}) and since
$1+s_{s\pm}>1$
\begin{eqnarray}\label{inqomegad2}
\lambda +\lambda^{\pi} < 0.5.
\end{eqnarray}

Let us to study now the conditions imposed on the found solutions
if we require stability. Evaluating the elements of the matrix
$\mathbf{M}$ at the critical points $(r_{s+},s_{s+})$ and
$(r_{s-},s_{s-})$, we obtain from the equation for the
eigenvalues,$det [\mathbf{M}-\eta \mathbf{I}]$, and since
$\partial g(r,s)/\partial r =0$, that
\begin{eqnarray}\label{eqeingenvalue}
\left [\left (\frac{\partial f(r,s)}{\partial r}\right
)_{|_{r=r_{s\pm},s=s_{s\pm}}} - \eta\right ]\left [\left
(\frac{\partial g(r,s)}{\partial s}\right
)_{|_{r=r_{s\pm},s=s_{s\pm}}}-\eta\right ]=0.
\end{eqnarray}
The eigenvalues are then
\begin{eqnarray}\label{eingenvalue1}
\eta_{1}=  \left( \frac{\partial f(r,s)}{\partial r}\right)
_{|_{r=r_{s\pm},s=s_{s\pm}}}= \,\,(\lambda
+\lambda^{\pi})(1+s_{s\pm}) +\omega_{D}-\omega_{m},
\end{eqnarray}
and
\begin{eqnarray}\label{eingenvalue2}
\eta_{2}=\left (\frac{\partial g(r,s)}{\partial s}\right
)_{|_{r=r_{s\pm},s=s_{s\pm}}}=\,\,(\lambda
+2\lambda^{\pi})(1+s_{s\pm}) +\lambda s_{s\pm}+
\omega_{D}-\omega_{X}.
\end{eqnarray}
The condition of stability, $\eta_{1}<0$ and $\eta_{2}<0$, implies
that the following constraints must hold
\begin{eqnarray}\label{constrat3}
\omega_{D}-\omega_{m}< -( \lambda +\lambda^{\pi})(1+s_{s\pm}),
\end{eqnarray}
and
\begin{eqnarray}\label{const3}
\omega_{D}-\omega_{X}< -\left[(\lambda
+2\lambda^{\pi})(1+s_{s\pm}) +\lambda s_{s\pm}\right ].
\end{eqnarray}
Notice that the constraint given by Eq.(\ref{constrat3}) is the
same obtained in Eq.(\ref{inqrsol}). Nevertheless, we need to look
for the range of $\omega_{D}-\omega_{X}$ which can accommodate the
constraint given by Eq.(\ref{inqforomegas}) and Eq.(\ref{const3}).
Choosing Eq.(\ref{const3}) as the constraint for the upper limit
of $\omega_{D}-\omega_{X}$, the upper limit indicated in
Eq.(\ref{inqforomegas}) is also satisfied.  We can impose the
condition
\begin{eqnarray}\label{XX}
- \left( (\lambda +2\lambda^{\pi})+2\sqrt{\lambda^{\pi}(\lambda
+\lambda^{\pi})}\right )< -\left[(\lambda
+2\lambda^{\pi})(1+s_{s\pm}) +\lambda s_{s\pm}\right ],
\end{eqnarray}
which leads to the following condition for $s_{s\pm}$
\begin{eqnarray}\label{XZ}
s_{s\pm} < \sqrt{\frac{\lambda^{\pi}}{\lambda +\lambda^{\pi}}}.
\end{eqnarray}
Since it is physically reasonable assume $s_{s\pm}\ll 1$ for a
late time evolution,

It is straightforward to check from the expressions
Eq.(\ref{constrat3}) and Eq.(\ref{const3}) that, independently of
the critical point considered, the equation of state of the non
decaying fluids satisfy
\begin{eqnarray}\label{Xvsm}
\omega_{X}>\omega_{m}.
\end{eqnarray}
If $\omega_{m}=0$, which is the equation of state for the dark
matter fluid, the above result indicates us that the exigency of
stability for the stationary solutions of the evolution equations
(\ref{requationprime}) and (\ref{sequationprime}), imposes an
unknown interacting fluid with non null pressure.

\section{Discussion}

In the present investigation we have considered a cosmological
scenario where the dark energy is decaying into the dark matter
and another component of the universe, which we do not identify
explicitly. We have assumed that this three fluids have an
equation of state with $\omega$ constant. We have choose three
different coupling terms, analyzing the stationary solutions of
the evolution equation for parameters $r$ and $s$.

When the coupling is proportional to the dark energy only, we have
found that the conditions for the stationary solutions be positive
are the same of those to be stable.  For dark matter with
negligible pressure we obtain that the dark energy must
necessarily have an equation of state with $\omega_{D}< 0$.

When the coupling is only proportional to the DM energy density,
since in this case it is not possible to obtain stable solutions
for the three interacting fluids, if $\lambda >0$ and $
\lambda^{\pi}>0$, which guaranty that DE decaying in the other
fluids. Nevertheless, relaxing this condition and taking $\lambda
+ \lambda^{\pi}<0$ and $\lambda^{\pi}>0$ we obtain stationary
solutions which are stable. Notice that this physically correspond
to a DM decaying in DE and in the third fluid. It is interesting
to mention that the usual case of DE interacting only with DM,
which have been discussed for this coupling in~\cite{Guo}, showed
that the data slightly favored a DM decaying in DE and
$\omega_{D}<-1$. Unfortunately in our approach, we obtain that the
unknown fluid and DE have positive pressure, leading to a
decelerated expansion. As in the case of DE interacting only with
DM, there is no stable solutions for DE with negative pressure.

In the third case studied, the coupling consider a different fluid
from those of the dark sector, taking a term proportional to the
sum of the DE density and the third fluid density. We have found
two fixed points and from the constraints derived from the
condition of stability we have obtain that
$\omega_{X}>\omega_{m}$, which means an interacting third fluid
with positive pressure. This type of coupling can then accommodate
an scenario with radiation~\cite{Jamil}, unparticle~\cite{Chen} or
even some form of warm DM~\cite{WDM} as the third interacting
fluid.

%\begin{figure}[th]
%\includegraphics[width=5.0in,angle=0,clip=true]{fig_1}
%\caption{...} \label{figura1}
%\end{figure}
%\begin{figure}[th]
%\includegraphics[width=5.0in,angle=0,clip=true]{fig_2}
%\caption{...} \label{figura2}
%\end{figure}

\section{acknowledgements}
NC and SL acknowledge the hospitality of the Physics Department of
Universidad de La Frontera where part of this work was done. SL
acknowledges the hospitality of the Physics Department of
Universidad de Santiago de Chile. This work was supported from
DIUFRO DI08-0041, of Direcci\'on de Investigaci\'on y Desarrollo,
Universidad de La Frontera (FP) and DICYT 040831 CM, Universidad
de Santiago de Chile (NC).

\end{document}